\documentclass[aps,prd,eqsecnum,onecolumn,showpacs,nofootinbib,preprintnumbers,a4paper]{revtex4}

\usepackage{graphicx}
\usepackage{bm}

\newcommand{\Mbar}{{\,\overline{\vphantom{M}\,\,\,\,\,}\!\!\!\!\!\!M}}

\def\bea{\begin{eqnarray}}
\def\eea{\end{eqnarray}}
\def\be{\begin{equation}}
\def\ee{\end{equation}}

\begin{document}
\begin{flushright}
\parbox[t]{1.8in}{USTC-ICTS-10-08}
\end{flushright}

\title{A Matter Bounce By Means of Ghost Condensation}

\author{Chunshan Lin$^{1,2)}$, Robert H. Brandenberger$^{1)}$ and Laurence
Perreault Levasseur$^{1)}$}

\affiliation{1) Physics Department, McGill University, Montreal, QC, H3A 2T8, Canada}

\affiliation{2) Interdisciplinary Center of Theoretical Studies, University of Science and Technology of China, Hefei, Anhui 230026, China}

\begin{abstract}

Assuming the existence of a scalar field which undergoes ``ghost condensation" and which
has a suitably chosen potential, it is possible to obtain a non-singular bouncing cosmology
in the presence of regular matter and radiation. The potential for the ghost condensate field
can be chosen such that the cosmological bounce is stable against the presence of
anisotropic stress. Cosmological fluctuations on long wavelengths relevant to current
cosmological observations pass through the bounce unaffected by the new physics
which yields the bounce. Thus, this model allows for the realization of the ``matter bounce"
scenario, an alternative to inflationary cosmology for the generation of the observed
primordial fluctuations in which the inhomogeneities originate as quantum vacuum
perturbations which exit the Hubble radius in the matter-dominated phase of
contraction.
\end{abstract}

\maketitle

\section{introduction}

The inflationary universe scenario \cite{Guth,others} is the current paradigm
for early universe cosmology. While phenomenologically successful,
inflationary models face various conceptual problems (see e.g. \cite{RHBrev1}),
most notably the singularity problem \cite{Borde}. Hence, it has been
of great interest to construct nonsingular bouncing cosmologies (for a
recent review see e.g. \cite{Novello}).

It is possible that a nonsingular cosmological bounce is followed by a
period of inflation (see e.g. \cite{Yifu1} for an example). However, it is also of
interest to explore alternatives to inflation which could result from a bouncing
cosmology. In \cite{Wands,Fabio1} it was realized that quantum vacuum
fluctuations which exit the Hubble radius in a matter-dominated phase
of contraction obtain a scale-invariant spectrum of curvature fluctuations
on super-Hubble scales. Provided that on the large scales corresponding
to current cosmological observations the fluctuations pass through the
bounce unaffected by the new physics which yields the cosmological
bounce, then we have a mechanism alternative to inflation for producing
a scale-invariant spectrum of cosmological perturbations. This alternative
is called the ``matter bounce scenario" (see e.g. \cite{RHBrev2} for
introductory reviews). The matter bounce scenario can be distinguished
from simple inflationary models through the shape and amplitude of
the bispectrum, the three-point function of cosmological perturbations,
which has a special shape and an amplitude which is large compared
to what results in simple inflationary models \cite{Xue}.

From the Penrose-Hawking singularity theorems \cite{Hawking} it
is clear that new physics is required in order to obtain a cosmological
bounce. There have to be modifications either to the gravitational
action, or else matter which violates the usual energy conditions
has to be introduced. There have been various suggestions on how
to obtain a cosmological bounce by means of modifying gravity.
For example, the ``non-singular Universe" construction of
\cite{MBS} yields a cosmological bounce, as does the ghost-free
higher derivative gravity action of \cite{Biswas}. It is also possible
to obtain a bounce in the context of ``mirage cosmology" where
our space-time is a brane moving in a curved higher-dimensional
background space \cite{Saremi}. As was realized in \cite{HLbounce},
in the presence of spatial curvature a cosmological bounce also
occurs generically in Ho\v{r}ava-Lifshitz gravity \cite{HL}.

If we maintain General Relativity as the theory of space-time, then
a cosmological bounce can also occur if we suitably modify the
matter sector. A generic way of obtaining a bounce is to add
to the usual matter sector which obeys the standard energy conditions
a ``ghost" sector with opposite sign kinetic energy terms in the
matter Lagrangian. This class of models is called
``quintom" models \cite{quintom} and can lead to a bounce
\cite{quintombounce}. One way to realize  a quintom bounce
is by considering a scalar field $\varphi$ with standard kinetic
term and with mass $m$ and adding a ghost scalar field
$\tilde \varphi$ with mass $M \gg m$. This model was considered
in detail in \cite{Taotao}. A specific realization is in the context of
the scalar field sector of the Lee-Wick Standard Model \cite{LW},
as studied in \cite{Taotao2}. A serious challenge for all
models with a ghost field is the ghost instability problem
(see e.g. \cite{Cline}).

All analyses of the evolution of cosmological fluctuations through
non-singular bounces (e.g. \cite{ABB} in the case of the model
of \cite{Biswas}, \cite{Saremi} in the case of a mirage cosmology
bounce, \cite{Taotao,Taotao2} in the case of quintom models and
\cite{Xue2} in the case of the Ho\v{r}ava-Lifshitz bounce) indicate
that the spectrum of the curvature fluctuations does not change
during the bounce phase on length scales larger than the duration
of the bounce.

There are two serious problems for matter bounce scenarios. The
first is the fact that some bounces are unstable to the presence of
radiation, the reason being that the energy density of radiation
increases faster as the universe contracts compared to the
increase in the effective energy in the terms yielding the bounce.
In particular, this was shown  \cite{Johanna} to be a serious
problem in the case of the bounce in the Lee-Wick model, and
the problem will likely also be serious in other models in which
the ghost field yielding the bounce is a scalar field. In contrast,
models which are based on asymptotically free gravity models
such as \cite{MBS} and \cite{Biswas} will be free from this problem,
as is the bounce in Ho\v{r}ava-Lifshitz gravity.

An even more serious problem is the instability of the contracting
pre-bounce phase to the presence of anisotropies since the
energy density in anisotropies scales as $a^{-6}$ ($a(t)$ being
the cosmological scale factor) compared to $a^{-4}$ for
radiation \footnote{This leads to the BKL \cite{BKL} chaotic
behavior at the ``Big Crunch" singularity.}. This implies
that the cosmological bounce cannot be described in terms
of a homogeneous and isotropic background.

In this paper, we present a ghost condensation model in
which a cosmological bounce can be obtained. Since there
is no ghost in the perturbative spectrum of the theory,
the model is free from the ghost instability problem
of \cite{Cline}. The model is also
free from the two problems we just mentioned
above. The effective equation
of state of the ghost condensate is $w > 1$ and hence,
as studied in \cite{EkpnoBKL} in the context of the
Ekpyrotic scenario \cite{Ekp}, the anisotropies do not
come to dominate and the cosmological background remains
well described in terms of a homogeneous and isotropic
metric.

Ghost condensation was proposed \cite{ghost1} as a
theoretically consistent modification of gravity in the
infrared. It is based on the analog of the Higgs
mechanism in the kinetic sector of a scalar field
Lagrangian. The kinetic term $X$ of the scalar
field $\phi$, i.e.
\be
X \, \equiv \, -g^{\mu\nu}\partial_{\mu}\phi\partial_{\nu}\phi \,
\ee
where we are choosing the metric signature to be (-, + + +),
appears in the Lagrangian in terms of a function $P(X)$
with a nontrivial minimum, such that the leading term in $X$
in the Lagrangian has the ``wrong" sign, i.e. appears
as a ghost. If the function $P(X)$ has a non-trivial minimum
at $X = c^2$ (the ``ghost condensate")
and the field configuration takes on this value,
then fluctuations about this configuration have normal
kinetic term. Thus, there is no ghost in the spectrum of
fluctuations about the ghost condensate.

It was soon realized that, although initially introduced \cite{ghost1}
to provide a new explanation for infrared effects in gravity, the
ghost-condensation mechanism has many applications in
cosmology at ultraviolet scales. For example, it can be used
\cite{ghost2} to provide a new mechanism for inflation. It
was realized in \cite{ghost3} that the ghost condensate mechanism
can also provide stable violations of the null energy condition
and thus used to construct non-standard cosmologies, including
bouncing ones. Ghost condensation was used in
\cite{ghost4,ghost5} to provide non-singular versions of
Ekpyrotic cosmology. In this paper we make use of the ghost
condensation mechanism to construct a non-singular matter
bounce.

The outline of the paper is as follows. In the following section we
briefly review the idea of ghost condensation. In Section 3 we
work out the requirements for the ghost field potential $V(\phi)$
in order to obtain a bounce, in particular a bounce which is
stable against the addition of radiation and anisotropic stress.
In Section 4 we discuss the transfer of cosmological fluctuations
through the ghost bounce phase. The final section contains
our conclusions and a discussion of the results.

To set the notation, we work in terms of a spatially flat
Friedmann-Robertson-Walker background with metric
\be
ds^2 \, = \, - dt^2 + a(t)^2 (dx^2 + dy^2 + dz^2) \, ,
\ee
where $a(t)$ is the scale factor, $t$ is the physical
time and $x, y$ and $z$ are the co-moving spatial
coordinates. The Hubble expansion
rate is $H(t) = \dot{a} / a$.

\section{Ghost Condensation}

The Lagrangian of ghost condensation takes the following general
form
\be \label{ghostLag}
\mathcal{L} \, = \, M^4P(X) - V(\phi) \, ,
\ee
where $M$ is the characteristic mass scale of ghost condensation
and we follow the usual convention in the ghost condensation literature
to take $\phi$ to have dimensions of length such that $X$ is
dimensionless. To obtain a ghost condensate, the
function $P(X)$ must have a non-trivial minimum. To avoid a
cosmological constant problem, we set the value of $P$ at
the minimum to be zero. We will use the prototypical example
\be \label{ghostP}
P(X) \, = \, \frac{1}{8}(X-c^2)^2 \, ,
\ee
where $c$ is a dimensionless
constant which has nothing to do with the speed of
light (we are following the notation of the original papers on ghost
condensation).
The ghost condensate necessarily breaks Lorentz invariance. The
homogeneous and isotropic ground state
corresponds to the background scalar field configuration
\be
\phi \, = \, c t \, .
\ee

The equation of motion of the ghost condensation field is
\be \label{ghostEOM}
(P' + 2 \dot{\phi}^2P'')\ddot{\phi} + 3HP'\dot{\phi} \, =
\, - M^{-4} \frac{\partial V}{\partial\phi}~,
\ee
where $P' \equiv \frac{\partial P}{\partial X}$.

The analogs of the Friedmann equations obtained by
coupling the homogeneous and isotropic ghost condensate
to Einstein gravity are
\be \label{FRW1}
3 M_p^2 H^2 \, = \, M^4 \bigl( 2 X P' - P \bigr) + V + \rho_m \, ,
\ee
and
\be \label{FRW2} 2M_p^2 \dot{H} \, = \, - 2 M^4 X P' - (1 + w_m)
\rho_m \, . \ee 
We have assumed that in addition to the ghost condensate, there
is also regular matter with energy density $\rho_m$ and
equation of state $w = w_m$, where $w$ is the ratio of
pressure density to energy density.

Let us now consider homogeneous fluctuations $\pi$ about the
ghost condensate field, i.e.
\be \label{ghostfluct}
\phi(t) \, = \, c t + \pi(t) \, .
\ee
Inserting into  (\ref{FRW1}) and (\ref{FRW2}) we obtain
the following expressions for the energy and pressure
densities of the homogeneous $\pi$ field:
\bea
\rho_{X} \, &=& \, M^4 c^3 \dot{\pi} \bigl( 1 + {\cal{O}} (\frac{\dot{\pi}}{c}) \bigr) + V \, ,
\label{rhofluct} \\
\rho_{X} + p_{X} \, &=& \, M^4  c^3 \dot{\pi}  \bigl( 1 + {\cal{O}} (\frac{\dot{\pi}}{c}) \bigr)  \, .
\label{pfluct}
\eea
In order to obtain a cosmological bounce, it is necessary to cancel
the positive energy density of regular matter with the negative energy
density of the ghost condensate. This can be achieved by
having negative ${\dot{\pi}}$.  To complete the system of equations
for the fluctuations of the ghost condensate, we write down the
variational equation with respect to $\phi$. To leading order in
 ${\cal{O}} (\frac{\dot{\pi}}{c})$ it is
 \be \label{fluctEOM}
 c^2 a^{-3} \partial_t \bigl( a^3 {\dot{\pi}} \bigr) \, = \, - 2 M^{-4} \frac{\partial V}{\partial \phi} \, .
 \ee

By inserting the ansatz (\ref{ghostfluct}) into the Lagrangian
(\ref{ghostLag}) it follows that
$P' + 2 \dot{\phi}^2P''\gg 0$ is the necessary condition for the
theory expanded about the condensate to be ghost
free.

\section{Bounce Induced by a Ghost Condensate}

In the absence of regular matter, a ghost condensate leads to a
cosmological bounce if the ghost field potential energy (which we
assume to be positive) is cancelled by the negative ghost
condensate energy. To leading order in ${\cal{O}} (\frac{\dot{\pi}}{c})$
the condition for the bounce point is
\be
M^4 c^3 \dot{\pi} \, = \, -V \, .
\ee
The second Friedmann equation, i.e. (\ref{FRW2}), then implies that
$\dot{H} > 0$ at the bounce point, i.e. the scale factor indeed makes
a transition from a contracting phase to an expanding phase.

To obtain a bounce, it is necessary to have a non-trivial potential
$V(\phi)$ since in the absence of a potential the equation of motion
(\ref{fluctEOM}) implies
\be \label{fluctEOM2}
a^3 \dot{\pi} \, = \, \rm{const}
\ee
and hence the bounce condition, which in the absence of a potential
is $\dot{\pi} = 0$, cannot be reached.

To obtain a matter bounce, we must consider the equations of motion
in the presence of regular matter. We assume that the universe
begins at very large negative times with the ghost condensate
in its ground state $X=c^2$ (or $\pi = 0$). The energy density
in regular matter increases in time as
\be
\rho_m(t) \, \sim \, a(t)^{-3(1 + w_m)} \, .
\ee
We are interested in the cases of cold matter ($w_m = 0$) and
radiation ($w_m = 1/3$), or in anisotropic stress for which $w_m = 1$.
From (\ref{fluctEOM2}) it follows that in the
absence of a potential $V$ for $\phi$
\be
\rho_{X} \, \sim \, a^{-3} \, .
\ee
Hence, the ghost energy density cannot catch up with the
matter energy density and no bounce is possible. Thus
we conclude that, as in the case without matter, also in the
presence of matter it is necessary to introduce a potential
$V(\phi)$ in order to obtain a bounce.

 In order to obtain a bounce, the energy density in the
 ghost field must increase sufficiently fast:
 \be
 \rho_X \, \sim \, a(t)^{-p} \, ,
 \ee
where $p > 3$ is the minimal requirement, $p > 4$ is required
if the bounce is to be stable against the presence of
regular relativistic radiation, and $p > 6$ is required
if the bounce is to be stable against the presence of
anisotropies.

In order for the ghost energy density to increase faster than
$a^{-3}$, we require
\be
\partial_t \bigl( a^3 \dot{\pi} \bigr) \, < \, 0
\ee
and hence from (\ref{fluctEOM})
\be
\frac{\partial V}{\partial \phi} \, > \, 0 \, .
\ee

We assume that the ghost field $\phi$ starts at large negative
values. Hence, we make the following ansatz for our
potential
\be \label{pot}
V(\phi) \, = \, V_0 M^{-\alpha} \phi^{-\alpha} \, ,
\ee
where $V_0$ is a constant with units of potential energy
density which sets the overall scale of
the ghost energy density, and $\alpha$ is a constant to be
determined by the requirement that the ghost energy
density increases sufficiently fast. This potential
diverges at $\phi = 0$. However, we know that
the ghost condensate Lagrangian should only be
trusted at energy scales lower than $M^4$. Hence,
we will assume that the divergence of the potential
is cut off when the potential reaches this value. For
example, we can assume that the potential tends
smoothly to the limiting value $M^4$ which is taken
on at $\phi = 0$ (see Fig. 1 for a sketch of this
potential).

\begin{figure}
\includegraphics[width=7cm]{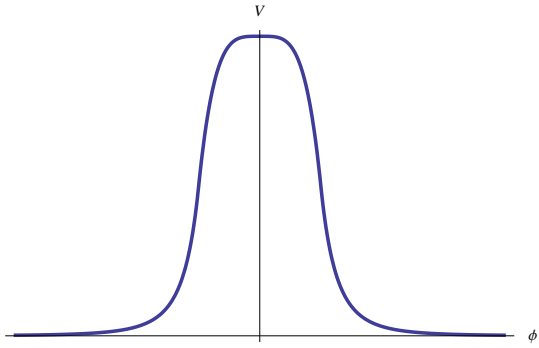}
\caption{Sketch of the effective potential of the ghost condensate field.}
\end{figure}

 Let us now derive the condition on $\alpha$ for which
 a non-singular bounce which is stable even in the
 presence of anisotropies will arise. For this, we
 return to the equation of motion (\ref{fluctEOM})
 which after inserting the ghost fluctuation ansatz
 (\ref{ghostfluct}) and expanding to leading order
 in $\dot{\pi}$ becomes
 \be
 c^2 \partial_t \bigl( a^3 \dot{\pi} \bigr) =
 - 2 a^3 M^{-4 - \alpha} \frac{\partial V}{\partial \phi} \, .
 \ee

 Inserting the form of the potential (\ref{pot}), the
 ansatz for the field $\phi$ from (\ref{ghostfluct}), and
 working in the range of times for which $|c t| \gg | \pi |$
 yields
 \be \label{fluctEOM3}
 \ddot{\pi} + 3 H \dot{\pi} \, =
 \, 2 c^{-2} V_0 M^{- 4 - \alpha} \alpha (ct)^{-(\alpha + 1)} \, .
 \ee
 We want to derive the condition on $\alpha$ such that in a
 background dominated by matter with an equation of state parameter
 $w$, the energy density of the ghost kinetic term which is proportional
 to $\dot{\pi}$ increases faster than the energy density in matter
 as the universe contracts. Hence, we insert the following
 form of the scale factor
 \be
 a(t) \, = \, (\frac{-t}{t_0})^{2/(3(1 + w))}
 \ee
 into (\ref{fluctEOM3}). We see that the source term (the right hand side
 of (\ref{fluctEOM3}) leads to
 \be
 \dot{\pi} \, \sim \, t^{-\alpha} \, .
 \ee
 and the condition for the ghost kinetic energy density to grow faster
 than the matter energy density is
 \be
 \alpha \, > \, 2 \, .
 \ee
 Since we wish the potential to be an even function of $\phi$, we can choose
 $\alpha = 4$. In this case the model is marginally stable against the
 addition of anisotropic stress. The case $\alpha = 6$ would provide
 a model which is stable.

Let us now summarize the evolution of the background cosmology in
our ghost bounce model. We begin at $t \rightarrow - \infty$ in a
contracting, matter-dominated phase with ghost condensate
field $\phi \rightarrow - \infty$ and the ghost condensate in
its ground state with $\pi = 0$. The energy density of the universe
is initially dominated by non-relativistic matter. As the universe
contracts, first radiation with $w = 1/3$ and then anisotropic
stress with $w = 1$ take over. However, the energy density
in the ghost condensate grows faster than any of these
energy densities and, at some point sufficiently close to $t = 0$
begins to dominate. Note that both the negative energy of
the ghost kinetic term and the positive potential energy of
the ghost field increase at the same rate. Eventually the
ghost field $\phi$ reaches the value when the potential
flattens out. At this point, the negative kinetic contribution to
the ghost energy catches up, and the bounce point $H = 0$ is
reached. Since at this point $\dot{\pi} < 0$ we have
$\dot{H} > 0$ and the universe begins to expand. The
ghost field evolves to positive values, and both the
ghost potential and kinetic energies begin to decrease rapidly,
allowing the energy densities of anisotropic stress, radiation
and cold matter to dominate again.

The evolution of the background has been studied numerically and
the results are presented in Figures  2 - 5. In all four figures,
the horizontal axis is time in Planck units. The vertical axis is
$\phi$, $\dot{\phi}$, and $H$ respectively. There is a nonsingular
bounce at time $t = 0$. The graphs are based on solving Eqs.
(\ref{ghostEOM}) and (\ref{FRW1}) using Mathematica, beginning the
evolution at the bounce point $t = 0$ where ``initial" conditions
(again in Planck units) $\phi(0) = 0$ and $\dot{\phi}(0) =
\sqrt{2/3}$ are chosen. The value of $M$ was taken to be $M = 2
\times 10^{-3}$, and the potential was taken to correspond to
$\alpha = 4$, with a value of $V_0$ such that 
\be
V(\phi) \, = \, \frac{1}{\phi^4 + 10^{12}} \, .
\ee
The energy density of matter sector was taken to be
$\rho_m(0)=10^{-12}$. As is obvious from the graphs, a smooth
nonsingular bounce results.

\begin{figure}
\includegraphics[width=7cm]{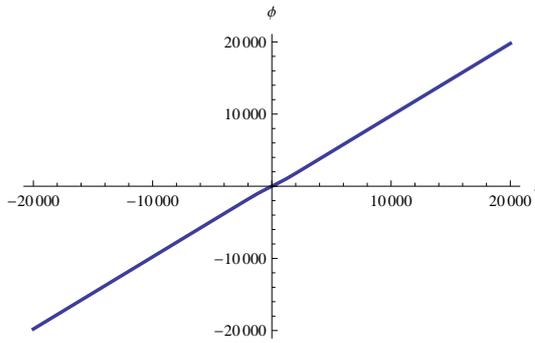}
\caption{Evolution of the ghost field $\phi$ as a function of
time in our numerical simulation.}
\end{figure}

\begin{figure}
\includegraphics[width=7cm]{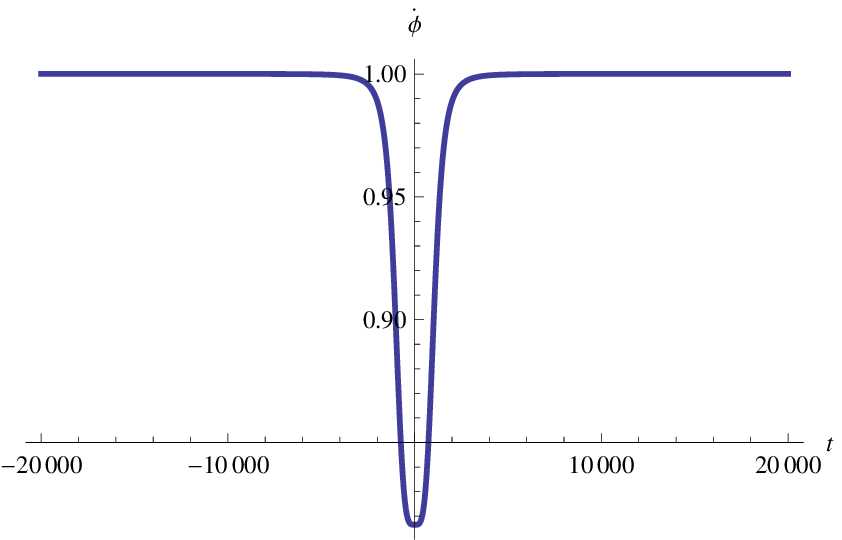}
\caption{Evolution of the ghost field velocity $\dot{\phi}$ as a function of
time in our numerical simulation.}
\end{figure}

\begin{figure}
\includegraphics[width=7cm]{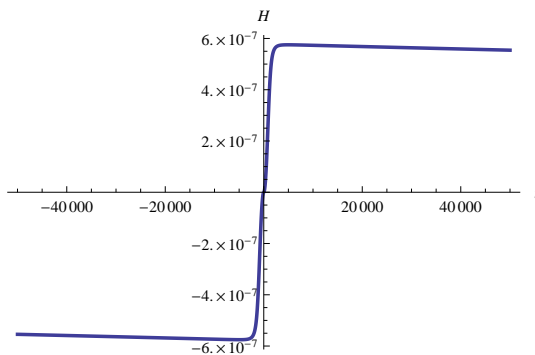}
\caption{Evolution of the Hubble parameter $H$ as a function of
time in our numerical simulation.}
\end{figure}


\section{A First Look at Cosmological Perturbations in the Ghost Condensate Model}

\subsection{Generating a Scale-Invariant Spectrum}

In this section, we will show that the perturbations
induced by the ghost condensate do not change the primordial spectrum of
curvature fluctuations induced by regular matter in the contracting
phase. Thus, in the matter bounce background the scale-invariant
spectrum of fluctuations on scales which exit the Hubble radius during
the mater-dominated phase of contraction will not be distorted by
the ghost condensate, neither in the contracting nor in the bounce
phase. First, in this subsection, we review how a scale-invariant
spectrum of curvature fluctuations arises in the matter bounce scenario.

To set the framework, we present in Figure 6 a space-time sketch
of a matter bounce. The horizontal axis is space, the vertical
axis is time, and the bounce time is $t = 0$. We are interested in
fluctuations on scales which are currently observable, and
assume that these scales exit the Hubble radius in a matter-dominated
phase of contraction. This assumption is not very
restrictive. If the bounce cosmology background is time-symmetric
and the expanding phase corresponds to our currently observed
universe, then the assumption is satisfied for all scales on which
the primordial matter power spectrum is currently well
measured \footnote{This background predicts a kink in the
matter power spectrum on scales which can be probed by
Lyman $\alpha$ observations - see \cite{LiHong} for a
discussion of this point.}. If the bounce is not time-symmetric,
then according to the Second Law of Thermodynamics
the expanding phase will have more entropy than the
initial collapsing one and the radiation phase is
shorter than in the case of the symmetric bounce, which means
that a wider range of scales exits the Hubble
radius in the matter-dominated phase.

 As shown in the Figure, scales of cosmological interest today
 exit the Hubble radius during the matter-dominated
 phase of contraction. If the bounce takes place at an
 energy scale comparable to that of Grand Unification, then the
 current Hubble radius corresponds to a physical length
 of about $1 {\rm{mm}}$ at the bounce. This scale is in
 the far infrared compared to the scale which sets the
 physics of the bounce.

\begin{figure}[htbp]
\includegraphics[scale=0.8]{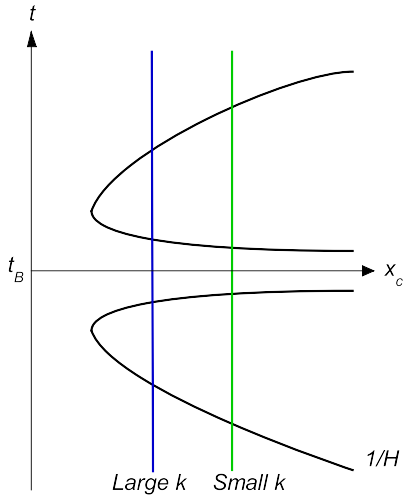}
\caption{A sketch of the evolution of perturbations with different
comoving wave numbers $k$ in the matter bounce. The horizontal
axis represents comoving coordinates, the vertical axis is time, with the
bounce point being $t = t_B$. The two vertical lines correspond to the
wavelengths of two different scales, the thick black curve gives the
Hubble radius.}
\label{fig:ghostbounce}
\end{figure}

The theory of cosmological perturbations in the model with both
regular matter and a ghost condensate in general FRW background
was worked out in \cite{Mukohyama:2006be}.  Our goal here is to
show that the presence of the ghost condensate and its induced
curvature fluctuations will not mess up the scale-invariant
spectrum of perturbations resulting from regular matter in a
matter-dominated phase of contraction. We have to show first that
the ghost fluctuations in the matter-dominated phase of
contraction do not grow faster than the fluctuations without the
presence of the ghost field. Secondly, we need to show that the
curvature fluctuations which are induced by the coupling between
regular matter and the ghost field are negligible, and thirdly, we
need to show that the spectrum of fluctuations on scales of
relevance to current cosmological observations is not changed
during the short phase where the ghost condensate dominated the
dynamics.

We begin with the general framework of metric fluctuations. Working in
in longitudinal gauge (see \cite{MFB} for a comprehensive
review of the theory of cosmological perturbations)  the
metric including scalar fluctuations take the form
\be
ds^2 \, = \, - (1 + 2 \Phi) dt^2 + a(t)^2 (1 - 2 \Psi) d{\bf x}^2 \, ,
\ee
where $\Phi$ and $\Psi$ are the two gravitational potentials which depend
on space and time and encode the metric perturbations. Since there
is no anisotropic stress at linear order (both in the regular matter sector
and in the ghost matter Lagrangian),
we have $\Psi = \Phi$ as a consequence of the perturbed off-diagonal
spatial Einstein equations. The remaining potential $\Phi$ is the
relativistic generalization of the Newtonian gravitational potential.

To describe fluctuations on super-Hubble scales, it is more
convenient to use the fluctuation variable $v$ in terms of which
the action for canonical fluctuations has canonical form \cite{Mukh}.
If matter is a simple scalar field $\varphi$, then $v$ can be expressed in
terms of the matter and metric fluctuations $\delta \varphi$
and $\Phi$ via
\be
v \, = \, a \bigl[ \delta \varphi + \frac{z}{a} \Phi \bigr] \,
\ee
where $z$ is a function of matter and metric background whose
form does not concern us here (see \cite{MFB} or \cite{RHBrev3}
for an introductory review). Closely associated with $v$ is
the variable ${\zeta}$, the curvature fluctuation in co-moving
gauge (called unitary gauge in the ghost condensate literature):
\be
\zeta \, = \, a^{-1} v \, .
\ee

In an expanding universe the dominant of the two modes of
$\zeta$ has constant amplitude. However, in a contracting
universe the dominant mode of $\zeta$ is increasing on
super-Hubble scales. In fact, in a matter-dominated phase
of contraction the dominant mode of $v$ scales as
(see \cite{Wands,Fabio1} for the original works and \cite{RHBrev2}
for recent reviews)
\be
v(t) \, \sim \, t^{-1/3} \, ,
\ee
and hence
\be \label{one}
\zeta(t) \, \sim \, t^{-1} \, .
\ee

The above is exactly the growth rate needed to convert an initial vacuum
spectrum for fluctuations on sub-Hubble scales into a scale-invariant
spectrum on super-Hubble scales. We briefly review the argument.
The vacuum spectrum for the canonical fluctuation variable states
that
\be \label{vacuum}
v(k, t) \, \sim \, k^{-1/2}
\ee
on sub-Hubble scales in the contracting phase. The condition for
the time $t_H(k)$ when the scale $k$ crosses the Hubble radius is
\be \label{crossing}
t_H(k) \, = \, k^{-1} a(t_H(k)) \, .
\ee
Making use of the scaling (\ref{one}), the vacuum initial conditions
(\ref{vacuum}) which hold until Hubble radius crossing, and
the Hubble radius crossing condition (\ref{crossing}) we
obtain
\be
\zeta(k, t) \, = \, \frac{t_H(k)}{t} a(t_H(k)) v(k, t_H(k)) \, \sim \, t_H(k)^{1/3} v(t_H(k))
\, \sim \, k^{-1} k^{-1/2} \, ,
\ee
which corresponds to a scale-invariant power spectrum. Note that in the
last step we have used the scaling $t_H(k) \sim k^{-3}$ which holds
in a matter-dominated phase of contraction.

Thus, we have shown that vacuum fluctuations which exit the Hubble radius
during a matter-dominated phase of contraction acquire a scale-invariant
spectrum of perturbations on super-Hubble scales. This spectrum is
maintained if the background equation of state changes when radiation
starts to dominate. We will have to show that the presence of a ghost condensate
does not change the evolution of the fluctuations in the contracting phase,
and that the spectrum is not
distorted in the phase dominated by a ghost condensate.

\subsection{The Evolution of Ghost Perturbations}

Let us now turn to the ghost matter sector. Introducing
a fluctuating field $\pi(t, {\bf x})$ which depends both on space and time,
the Lagrangian to quadratic order in $\pi$ about a non-trivial
background $\dot{\pi_0}$ (the $\dot{\pi}$ appearing in the previous
section) is \cite{ghost3}
\be \label{ghostLag2}
{\cal L} \, = \, \frac{1}{2} M^4 \dot{\pi}^2 + \dot{H} M_{pl}^2 (\nabla \pi)^2
- \frac{1}{2} {\tilde{M}}^2 (\nabla^2 \pi)^2 \, ,
\ee
where $\nabla$ is the spatial gradient operator. Note that to
leading order in $\pi$, no quartic order spatial gradient terms
appear if we only use the Lagrangian (\ref{ghostLag}), and to
obtain (\ref{ghostLag2}) we had to introduce higher derivative
terms, following \cite{ghost1,ghost3}. To derive the form of the
second term in (\ref{ghostLag2}), we made use of the ghost
background equations of the previous section.

By working in unitary gauge in which $\zeta$ appears directly
as the fluctuation of the diagonal term in the spatial metric, it was
shown in Section 3.3 of \cite{ghost3} that for long wavelength modes
\be \label{two}
{\dot \zeta} \, \sim \, a^{-3} \, \sim \, t^{-2} \, ,
\ee
where we have used the matter-dominated scaling of $a(t)$ in the last step.
By comparing (\ref{one}) and (\ref{two}) we see that the matter and
ghost-induced fluctuations grow at the same rate on large scales.

To show that the presence of a ghost condensate sector will not
change the dominant contribution to the spectrum of cosmological
fluctuations in the phase before the ghost condensate begins to
dominate, we need to study the spectrum of the metric
fluctuations induced by the presence of coupling between the
matter and ghost sector. To do this, we make use of the equations
for the metric fluctuation variable $\Phi$ which were derived in
\cite{Mukohyama:2006be}.

As matter we consider a ghost condensate in addition to
a matter fluid, and we focus on a matter-dominated contracting background.
Thus, the Einstein equations are
\begin{eqnarray}
M_{pl}^2G_{\mu\nu} \, = \, T^{(\phi)}_{\mu\nu}+T_{\mu\nu}~,
\end{eqnarray}
where we use a superscript $(\phi)$ to denote the ghost condensate part.
We split $\Phi$ into the contribution $\Phi_m$ which would be
obtained in the absence of matter fluid, and a term $\Phi_g$ which
is induced by the presence of the ghost:
\be
\Phi \, = \, \Phi_m + \Phi_g \, .
\ee
If we neglect the anisotropic stress and entropy perturbation, the
general equations of motion for $\Phi$ are \cite{Mukohyama:2006be}
\begin{eqnarray} \label{2eqsofNewt}
\partial_t \Phi_{g} + H \Phi_{g}  &=&  \frac{\alpha}{2} \frac{\mathbf{k}^2}{a^2} \chi \nonumber\\
\partial_t \chi  &=&
\bigl( \frac{M^2}{M_{pl}^2} - \frac{2}{M^2}\frac{\mathbf{k}^2}{a^2} \bigr) \Phi_{g} + S_{\chi} \, ,
\end{eqnarray}
where $\alpha = {\tilde M}^2/M^2$ and the source term is given by
\begin{eqnarray}
M_{pl}^2 \frac{\mathbf{k}^2}{a^2} S_{\chi}  &=&  \frac{M^2}{2M_{pl}^2}\delta\rho~, \nonumber\\
\frac{\mathbf{k}^2}{a^2}\Phi_{m}  &=&  \frac{\delta\rho}{2M_{pl}^2}~.
\end{eqnarray}
Note that the two equations in (\ref{2eqsofNewt})  come
from combining all of the perturbed Einstein equations, i.e. the
$0-0$, $0-i$, and $i-i$ components. Thus, the set
of these equations is complete. Combining these two equations we obtain
the following equation of motion for the Newtonian potential:
\begin{eqnarray}\label{eomNEWt}
\partial_{t}^{2}\Phi_{g} + 3H\partial_{t}\Phi_{g} + (2H^{2}+\dot{H})\Phi_{g} + \frac{\alpha}{M^{2}}\left(\frac{\mathbf{k}^{2}}{a^{2}}\right)^{2}\Phi_{g} - \frac{\alpha M^{2}}{2M_{pl}^2}\frac{\mathbf{k}^{2}}{a^{2}}\Phi_{g} \, & = & \,
\frac{\alpha}{2}\frac{M^{2}}{M_{pl}^{2}}\frac{\mathbf{k}^{2}}{a^{2}}\Phi_{m} \, .
\end{eqnarray}
This equation can be simplified by factoring out the expansion/contraction of
space via defining the new variable $\tilde{\Phi}=a(t)\Phi$, and by working with
conformal time $\tau$ defined via $dt = a(t) d\tau$. Then, the equation
of motion for the Newtonian potential can be rewritten as
\begin{eqnarray} \label{ghosteq}
\tilde{\Phi}_{g}'' + \frac{81\alpha}{M^{2}} \frac{\mathbf{k}^{4}t_0^4}{\tau^{4}} \tilde{\Phi}_{g}   - \frac{\alpha M^{2}}{2M_{pl}^2}\mathbf{k}^{2}\tilde{\Phi}_{g}  & = &
\frac{\alpha}{2}\frac{M^{2}}{M_{pl}^{2}}\mathbf{k}^{2}\tilde{\Phi}_{m}~,
\end{eqnarray}
where a prime indicates the derivative with respect to $\tau$.
We have already used the matter contracting background
$a = \frac{t}{t_0}^{2/3}$ (where $t_0$ is some reference time)
to get the above equation.

We will solve (\ref{ghosteq}) in the large wavelength approximation in
which we neglect the $k^4$ term. Then, the general solution is a
combination of the general solution $\Phi_{g}^{h}$ of the homogeneous
equation and a particular solution $\Phi_{g}^{p}$ of the inhomogeneous equation
which can be determined by the Green function method. To simplify
the notation we introduce the constants
where
\begin{eqnarray}
\beta \, &\equiv& \, \frac{81\alpha}{M^2}~,\nonumber\\
\gamma \, &\equiv& \, \frac{\alpha M^2}{2 M_{pl}^2}
\end{eqnarray}
The general solution of the homogeneous
equation is
\be
\tilde{\Phi}_{g}^{h} \, = \, c_1 e^{k \sqrt{\gamma} \tau} + c_2 e^{- k \sqrt{\gamma} \tau} \, ,
\ee
where $c_1$ and $c_2$ are constants which multiply the two basis solutions
of the equation $C_1(\tau)$ and $C_2(\tau)$, respectively. The Wronskian
$\epsilon(\tau)$ obtained from the two basis solutions is
\be
\epsilon(\tau) \, = \, (2 k \sqrt{\gamma})^{-1}
\ee
in terms of which the particular solution $\Phi_{g}^{p}$ becomes
\be \label{phipart}
\tilde{\Phi}_{g}^{p}(\tau) \, = \, C_1(\tau) \int d\tau^{'} \epsilon(\tau^{'}) C_2(\tau^{'}) {\cal{S}}(\tau^{'})
-  C_2(\tau) \int d\tau^{'} \epsilon(\tau^{'}) C_1(\tau^{'}) {\cal{S}}(\tau^{'}) \, ,
\ee
where ${\cal{S}}(\tau)$ is the source term in (\ref{ghosteq}).
Recall that the general solution for
$\Phi_{m}$ on super Hubble scales \cite{Fabio1} is
\begin{eqnarray}
\tilde{\Phi}_{m} \, = \, a(t)\Phi_{m} \, = \, D\tau^2 + \frac{S}{\tau^3}~,
\end{eqnarray}
where $D$ and $S$ are constants. Inserting this equation
into the source term in (\ref{ghosteq}), and approximating
the mode functions $C_1(\tau)$ and $C_2(\tau)$ of the
homogeneous equations as constant (which is justified over
short time intervals for long wavelength modes), we obtain
the following approximate form for the particular solution
\be \tilde{\Phi}_{g}^{p}(\tau) \, \simeq \, k^2\gamma\tau\bigl(
\frac{D}{3} \tau^3 - \frac{S}{2} \tau^{-2} \bigr) \, . \ee 

Note that the effects of the $k^4$ term in the equation of motion
(\ref{ghosteq}) can be included to leading order using the Born
approximation. In this approximation, the contribution to
$\tilde{\Phi}_{g}$ induced by the $k^4$ term is obtained by taking
the $k^4$ term to the right-hand side of (\ref{ghosteq}),
evaluating it for the homogeneous solution $\tilde{\Phi}_{g}^{h}$,
and using it as a second source term in (\ref{phipart}) to obtain
a second contribution to the particular solution. This source term
scales as $\tau^{-4}$, and hence its contribution to
$\tilde{\Phi}_{g}^{p}$ will scale as $\tau^{-2}$, it also grows
slower than $\tilde{\Phi}_{m}$. On the other hand, the spectrum is
suppressed by a factor of $k^3$ and is thus highly blue and
completely negligible on scales relevant to current cosmological
observations.

Returning to the contribution to $\tilde{\Phi}_{g}^{p}$
which dominates on large scales, we conclude that on
super-Hubble scales the leading term comes from $\tilde{\Phi}_{g}^{p}$
and is
\begin{eqnarray}
\Phi_{g} \, \sim \, a(t)^{-1} \tilde{\Phi}_{g} \, \sim \,
\frac{k^2 \gamma S}{\tau^3}.
\end{eqnarray}
Comparing with the matter perturbation
\be
\Phi_{m} \sim \frac{S}{\tau^5} \, ,
\ee
we see that the perturbation spectrum of the terms generated by the
ghost condensate is blue,  and that it grows more slowly than the
contribution from the matter sector.
This is a self consistent result. As analyzed at the beginning of
this section, the spectrum of the perturbations induced by matter is
scale-invariant because the amplitude of the perturbations which
cross the Hubble radius at an earlier time grows and catches up to the amplitude
of the perturbations which cross the Hubble radius at a later time in the
contracting phase. Since the perturbation induced by the ghost condensate
grows slower than that induced by matter, the perturbations which
cross the Hubble radius at an earlier time  cannot
catch up to the amplitude of the perturbations which cross the Hubble radius
at a later time, and thus the spectrum is blue. The scale-invariant
spectrum of the perturbations induced by matter will be preserved in the
contracting phase.

The third step is to show that the spectrum remains unchanged
around the bounce phase. We only need to show that the amplitude
of the curvature fluctuation induced by the ghost condensate will
not grow too much during the bounce phae. This can be seen
easily from Eq.(\ref{eomNEWt}).
Since the duration of
bounce phase is very short and scale factor $a(t)$ is almost a
constant we can parameterize the background as follows:
\begin{eqnarray}
H \, = \, \theta\cdot(t-t_B)~,
\end{eqnarray}
where $t_B$ is the time at the bounce point, and $\theta$ is a constant
with $\theta \gg H_c^2$, where $H_c$ is the value of $H$ at
the matching surface between the contracting phase and the bounce phase
\footnote{Note that this implies that the duration $\delta t$ of the bounce phase
satisfies $\sqrt{\theta} \delta t \ll 1$.} .
On scales of interest to current observations we have
$\frac{k^2}{a^2} \ll H_c^2 \ll \theta$, and thus Eq.(\ref{eomNEWt}) can be rewritten as
\begin{eqnarray}\label{eombphase}
\partial_t^2\Phi_g + \theta\Phi_g \, = \, 0~,
\end{eqnarray}
the solution of which is
\be
\Phi_g \, = \,  d_1 e^{i\sqrt{\theta}t} + d_2 e^{-i\sqrt{\theta}t} \, .
\ee
Since the bounce phase is so short we thus have
\be
\Phi_{g}(t) \, \simeq \, \Phi_g^c
\ee
also at the end of the bounce phase (where $\Phi_g^c$ is the
Newtonian potential induced by the ghost condensate at the beginning of
the bounce phase). The perturbation induced by the ghost is almost the
same as at the end of the contracting phase.

Note that near the bounce point the perturbation re-enters the Hubble radius
for a very brief time interval,  but $k^2\ll \theta$ is still true since scale factor
is almost a constant during the bounce phase,
and thus Eq. (\ref{eombphase}) is still valid
for describing the perturbations near bounce point.  We thus see that, as
long as the perturbations induced by the ghost are unimportant during the
contracting phase, they will also be unimportant during the bounce
phase.

Previous work in the context of a wide range of non-singular bouncing
models (see e.g. \cite{Saremi, Taotao, Taotao2, ABB, Xue2})
has shown that the matter-induced fluctuations
do not change during the bounce phase. Thus, we conclude
that the full spectrum of cosmological perturbations in our model
does not change during the bounce phase.

\subsection{Stability of Ghost Perturbation}

As derived in Section 3.3 of \cite{ghost3}, the action for the curvature
fluctuation variable $\zeta$ is
\be \label{zetaaction}
S \, = \, \int d^3 x dt a^3(t) \bigl[ A(t) {\dot{\zeta}}^2 + B(t) ( \frac{\nabla \zeta}{a})^2
+ C(t)  ( \frac{\nabla^2 \zeta}{a^2})^2 \bigr] \, ,
\ee
where the coefficient functions $A(t), B(t)$ and $C(t)$ are given in
the Appendix. Near the bounce point we can take the scale factor
to be a constant and neglect the terms proportional to $H$ which
appear in the  coefficient functions. Then, for each Fourier mode
we obtain a harmonic oscillator equation, and the dispersion relation
is
\be \label{disp}
\omega^2 \, = \, \frac{ - (\tilde{M}^2 M^4 + 4 M_{pl}^4 {\dot{H}})k^2
+ 2 M_{pl}^2 {\tilde{M}}^2 k^4}{2 M_{pl}^2 M^4} \, .
\ee

As is obvious from (\ref{disp}), there is an instability for all long wavelength
modes. In fact, the instability is a combination of two instabilities which
can be seen \cite{ghost1,ghost3} at a heuristic level,
the first being a gradient
instability which appears if $\dot{H} > 0$ as it is near the bounce,
and the second a Jeans type instability. These are due to the
second and first terms, respectively, in the coefficient of the $k^2$
term in (\ref{disp}). It is easy to determine the value of $k$ for which
the instability is the strongest. The exponent $\omega_c$ of the
instability is
\be
\omega_c \, = \, \frac{1}{4} \frac{\tilde{M} M^2}{M_{pl}^2}
+ \dot{H} \frac{M_{pl}^2}{M^2 \tilde{M}} \, .
\ee

We will now argue that the bounce period is too short for this instability
to change the spectrum of cosmological perturbations. To
estimate the time period $\Delta t$ of the bounce, we use
\be
\Delta t \dot{\phi} \, = \, \Delta \phi \, ,
\ee
where $\Delta \phi$ is the interval of $\phi$ corresponding to the
bounce. We will estimate this distance by the field value
for which the potential (\ref{pot}) approaches its maximal value
which we take to be $V(\phi) = M^4$. This gives
\be
\Delta \phi \, \sim \, M^{-1} \bigl( \frac{V_0}{M^4} \bigr)^{1/\alpha} \, .
\ee
Inserting $\dot{\phi} \simeq c$ and setting the constant $c = 1$
we obtain
\be \label{deltaomega}
\Delta t \omega_c \, \sim \,  \bigl( \frac{V_0}{M^4} \bigr)^{1/\alpha}
\bigl[ \frac{1}{4} \frac{\tilde{M} M}{M_{pl}^2}
+ \dot{H} \frac{M_{pl}^2}{M^3 \tilde{M}} \bigr] \, .
\ee
The first term in the square bracket is clearly much smaller than one.
To estimate the value of the second term, we make use of
(\ref{FRW2}) and (\ref{pfluct}) to estimate the value of $\dot{H}$:
\be
\dot{H} \, \sim \, \frac{M^4 \dot{\pi}}{M_{pl}^2} \,
\ee
with which the second term in the square bracket of (\ref{deltaomega})
is of the order of $\dot{\pi}$ which cannot be larger than order unity.
Thus, if $V_0 \ll M^4$ then
\be
\Delta t \omega_c \, \ll \, 1 \, ,
\ee
and the instability in the ghost condensate phase does not have time
to develop.

\section{Discussion and Conclusions}

In this paper we have presented a realization of the ``matter bounce"
by means of a ghost condensate. Compared to other realizations
of the matter bounce using a modification of the matter sector,
our model has several advantages: first of all, there is no ghost
in the perturbative spectrum of states. Secondly, the background
cosmology is stable against the addition of both regular radiation
and of anisotropic stress, the latter implying that our model
will likely be free of the chaotic mixmaster behavior which
plagues many models.

We have also studied the evolution of cosmological
perturbations through the bounce. We considered metric
fluctuations on super-Hubble scales in a model with both
background matter (a perfect fluid with equation of state
$p = 0$) and a ghost condensate. We have shown that
the ghost-induced fluctuations grow less fast during the
contracting phase than the curvature perturbations in a pure
matter model, and that in addition they have a blue
spectrum. Hence
the scale-invariance of curvature fluctuations on
super-Hubble scales in the contracting phase (assuming
that the inhomogeneities originate as quantum
vacuum fluctuations) is maintained. Since
the curvature perturbations which are induced by the
gravitational coupling between ghost condensate and
regular matter have a blue spectrum
they are irrelevant on scales of interest to current
cosmological observations. Finally, we have studied the
evolution of the fluctuations through the bounce phase,
a phase dominated by the ghost condensate. There
is a gradient instability, but the bounce phase is
sufficiently short such that this instability has no
time to develop.

In conclusion, we have shown that our ghost condensate background
cosmology provides a realization of the matter bounce
alternative to inflation for generating a scale-invariant
spectrum of adiabatic fluctuations.

\begin{acknowledgments}

This work is supported in part by a NSERC Discovery Grant, by a
grant from the FQXi, and by a Killam Research Fellowship awarded
to R.B. We would like to thank Shinji Mukohyama, Yi Wang, and Wei Xue
for useful discussions.

\end{acknowledgments}

 \appendix
 \section{Jeans instability during late time evolution}

In this subsection we discuss the Jeans instability during late
time evolution. The scalar excitation of the ghost condensate 
gives rise to an instability of the vacuum analogous to the Jeans
instability of pressureless matter coupled to gravity. The length
and time scales associated with the Jeans instability are
\begin{eqnarray}
L_J \, \sim \, \frac{M_p}{M^2},~~~T_J \, \sim \, \frac{M_p^2}{M^3}.
\end{eqnarray}
Any fluctuation with length scale larger than the Jeans length will be
unstable. Jeans collapse produces lots of lumps of scalar
excitation, and the universe will be filled with lumps. These
lumps may produce phenomenona which are 
inconsistent with current
cosmological observations, like excess lensing of light, 
supernova time delays and so on. These effects increase
in magnitude as $M$ increases. If matter is dominated by
the ghost condensate field as will be the case if the ghost
condensate is used to model dark energy, 
an upper bound on the energy scale $M$ 
of ghost condensation has been derived in Ref. \cite{ArkaniHamed:2005gu}:
\begin{eqnarray}
M \, < \, 100{\rm GeV} \, .
\end{eqnarray}

However, in our model in which the ghost field potential term
plays an important role, the ghost condensate only dominates
matter around the bounce point. Thus, our model is
free from the above constraint. For example, if we set $\alpha=4$ which
corresponds to the ghost condensate being marginally stable against
anisotropic stress, then the energy density of the ghost condensate 
scales as $a^{-6}$, and will be diluted away rapidly at late times in
the post-bounce phase. It is a standard result in the theory of
cosmological perturbations (see e.g. \cite{Peebles, Paddy} for
textbook discussions) that fluctuations in a subdominant
component are held back and cannot grow more than
logarithmically on scales where the fluctuations in the dominant
component are stable. For example, matter fluctuations in the
radiation epoch of Standard Cosmology only grow logarithmically
on sub-horizon scales. This implies that the ghost condensate does
not clump in the radiation phase of Standard Cosmology, and that
the ghost fluctuations track the usual matter fluctuations in the
matter epoch. Hence, in our model there are no constraints from late time
cosmology on the scale $M$.

 \section{Various Coefficients}

 In this subsection we give the expressions for the coefficient functions
 $A(t), B(t)$ and $C(t)$ which appear in (\ref{zetaaction}). They are
 taken from Section 3.3 of \cite{ghost3}:
\be \label{eq:A} A(t) = \frac{2\,M_{\rm Pl}^4\,\left(M^4 - 9
\Mbar^2\,H^2 - 2\,M_{\rm Pl}^2\, \dot H \right) }
  {4\,M_{\rm Pl}^4\,H^2 + \Mbar^2  M^4 }
\ee
\begin{eqnarray}
B(t) =\frac{M_{\rm Pl}^2}{(4 M_{\rm Pl}^4 H^2 + \Mbar^2 M^4)^2}
\left [-24 M_{\rm Pl}^6 \Mbar^2 H^4 \right. \!\!\! & + & \!\!\!
\Mbar^2 (M^4 -2 M_{\rm Pl}^2 \dot H) (M^4 \Mbar^2 - 4 M_{\rm Pl}^4
\dot H) \nonumber \\  + \; 4 H^2 (M^4 M_{\rm Pl}^4 \Mbar^2 + 4
M_{\rm Pl}^8 \dot H) \!\!\! & - & \!\!\! \left. 8 M_{\rm Pl}^6
\Mbar^2 H \ddot H \right]
\end{eqnarray}
\be C(t) = - \frac{2 M_{\rm Pl}^4 \Mbar^2}{4 M_{\rm Pl}^4 H^2 +
\Mbar^2 M^4}\;. \ee

\end{document}